# Recipes for Translating Big Data Machine Reading to Executable Cellular Signaling Models

Khaled Sayed[1], Cheryl A. Telmer[2], Adam A. Butchy[3], and
Natasa Miskov-Zivanov[1,3,4]

[1]Department of Electrical and Computer Engineering, University of Pittsburgh, PA, USA
[2]Department of Biological Sciences, Carnegie Mellon University, Pittsburgh, PA, USA
[3]Department of Bioengineering, University of Pittsburgh, Pittsburgh, PA, USA
[4]Department of Computational and Systems Biology, University of Pittsburgh, PA, USA
```
k.sayed@pitt.edu, ctelmer@cmu.edu, aab133@pitt.edu,
                    nmzivanov@pitt.edu
```

**Abstract.** With the tremendous increase in the amount of biological literature, developing automated methods for extracting big data from papers, building models and explaining big mechanisms becomes a necessity. We describe here our approach to translating machine reading outputs, obtained by reading biological signaling literature, to discrete models of cellular networks. We use outputs from three different reading engines, and describe our approach to translating their different features, using examples from reading cancer literature. We also outline several issues that still arise when assembling cellular network models from state-of-the-art reading engines. Finally, we illustrate the details of our approach with a case study in pancreatic cancer.

## 1 Introduction

Biological knowledge is voluminous; it is nearly impossible to read all scientific papers on a single topic such as cancer. When building a model of a particular biological system, one example being cancer microenvironment, researchers usually start by searching for existing relevant models and by looking for information about system components and their interactions in published literature.

Although there have been attempts to automate the process of model building [1, 2], most often modelers conduct these steps manually, with multiple iterations between (i) information extraction, (ii) model assembly, (iii) model analysis, and (iv) model validation through comparison with most recently published results. To allow for rapidly modeling the complexity of diseases like cancer, and for efficiently using ever-increasing amount of information in published work, we need representation standards and interfaces such that these tasks can be automated. This, in turn, will allow researchers to ask informed, interesting questions that can improve our understanding of health and disease.

The systems biology community has designed and proposed a standardized language for representing biological models is the *systems biology markup language* (SBML), which allows for using different software tools without the need for recreating models specific for each tool and allows also for sharing the built models between the different research groups [3]. However, the SBML standard is not easily understood by biologists who create mechanistic models. Therefore, software tools have been developed to provide biologists with an interface that allows them to focus on the modeling tasks by hiding the details of the SBML language [4-7].

To this end, the contributions of the work presented in this paper include:

• A *representation format* that is straightforward to use by both machines and humans, and allows for efficient synthesis of models from big data in literature.

• An *approach to effectively use* state-of-the-art machine reading output to create executable discrete models of cellular signaling.

- A *proposal for directions* to further improve automation of assembly of models from big data in literature.

In Section 2, we briefly describe cellular networks, our modeling approach, and our framework that integrates machine reading, model assembly and model analysis. In Section 3, we present details of our model representation format, while Section 4 outlines our approach to translate reading output to the model representation format. Section 0 discusses other issues that need to be taken into account when building interface between big data reading and model assembly in biology. Section 6 describes a case study that uses our translation methodology. Section 7 concludes the paper.

## 2   Background

### 2.1   Cellular networks

*Intra-cellular networks* include signal transduction, gene regulation and metabolic networks [8]. Signaling networks are characterized by protein phosphorylation and binding events, which transduce extracellular signals across the plasma membrane and through the cytoplasm [9]. Gene regulatory networks involve translocation of signaling proteins from the cytoplasm to the nucleus, where the integration of these protein signals act on the genome, resulting in changes in gene expression and cellular processes [10]. The regulation of metabolic networks incorporates phosphorylation and binding, as do signaling networks, and also integrates allosteric regulation, other protein modifications, and subcellular compartmentalization [11].

*Inter-cellular networks* assume interactions between cells of the same or different types. These interactions occur via signaling molecules such as growth factors and cytokines, synthesized and secreted by one cell, and bound to itself or other cells in its surrounding, or via a cell-cell contact.

At all levels of signaling, there are feedforward and feedback loops and crosstalk between signaling pathways to either maintain homeostasis or amplify changes initiated by extracellular signals [12].

### 2.2   Modeling approach

When generating executable models, we use a discrete modeling approach previously described in [13] when developing our models. As illustrated in the example in Figure 1, we represent system components as model *elements* (A, B, and C in the example), where each element is defined as having a discrete number of levels of activity. Next, each element has a list of regulators called *influence set*. In our example, A is a positive regulator of C, B and C are positive regulators of A, and C activates itself while B inhibits itself. Finally, each element has a corresponding *update rule*, a discrete function of its regulators. In our example, A is a conjunction of B and C, while C is a disjunction of A and C. Although the model structure is fixed, the simulator that we use [14] is stochastic, and allows for an realistic recapitulation of the behavior of system components and pathways in the network.

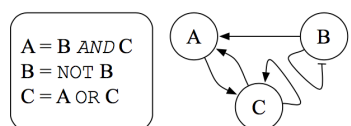

**Fig. 1.** Toy example illustrating our modeling approach.

### 2.3   Framework overview

To automatically incorporate new reading outputs into models, we have developed a reading-modeling-explanation framework, called DySE (<u>D</u>ynamic <u>S</u>ystem <u>E</u>xplanation), outlined in Figure 2. This framework allows for (*i*) expansion of existing models or assembly of new models from machine reading output, (*ii*) analysis and explanation of models, and (*iii*) generation of machine-readable feedback to reading engines. In this paper, we describe *the front end of the framework, the translation from reading outputs to the list of elements, and their influence sets, that accounts for available context information*.

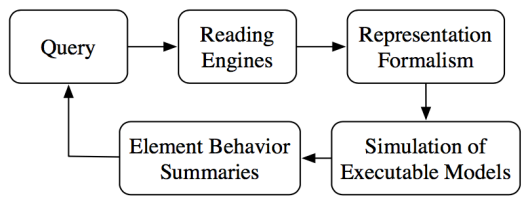

**Fig. 2**. DySE framework.

## 3   Model representation format

To enable comprehensive translation from reading engine outputs to executable models, the models are first represented in tabular format. It is important to note here that the tabular representation does not include final update rules, that is, the tabular version of the model is further translated into an *executable model* that can be simulated. Each row in the model table corresponds to one specific model element (i.e., modeled system component), and the columns are organized in several groups: (*i*) information about the modeled system component, (*ii*) information about the component's regulators, and (*iii*) information about knowledge sources. This format enables straightforward model extension to represent both additional system components as new rows in the table, and additional component-related features by including new columns in the table. The addition of new columns occurs with improvements in machine reading.

The first group of fields in our representation format includes **system component-related** information. This information is either used by the executable model, or kept as background information to provide specific details about the system component when creating a hypothesis or explaining outcomes of wet lab experiments.

A. `Name` – full name of element, e.g., "Epidermal growth factor receptor".
B. `Nomenclature ID` – name commonly used in the field for cellular components, e.g., "EGFR" is used for "Epidermal growth factor receptor".
C. `Type` – these are types of entities used by reading engines as listed in Table 1.

**Table 1**. Element type and ID database.

| Element Type | Database Name |
| --- | --- |
| Protein | [15] |
| Protein Family | [16], [17] |
| Protein Complex | [18] |
| Chemical | [19] |
| Gene | [20] |
| Biological process | [21], [22] |

**Table 2.** The list of cellular locations and their IDs from the Gene Ontology [21] database.

| Location Name | Location ID |
| --- | --- |
| Cytoplasm | GO:0005737 |
| Cytosol | GO:0005829 |
| Plasma Membrane | GO:0005886 |
| Nucleus | GO:0005634 |
| Mitochondria | GO:0005739 |
| Extracellular | GO:0005576 |
| Endoplasmic Reticulum | GO:0005783 |

D. `Unique ID` – we use identifiers corresponding to elements that are listed in databases, according to Table 1.
E. `Location` - we include subcellular locations and the extracellular space, as listed in Table 2.
F. `Location identifier` – we use location identifiers as listed in Table 2.
G. `Cell line` – obtained from reading output.
H. `Cell type` – obtained from reading outputs
I. `Tissue type` – obtained from reading output.
J. `Organism` – obtained from reading output.
K. `Executable model variable` – variable names currently include above described fields B, C, E, and H.

The second group of fields in our representation includes **component regulators-related** information that is mainly used by executable models, with a few fields used for bookkeeping, similar to the first group of fields.

L. `Positive regulator nomenclature IDs` – list of positive regulators of the element.
M. `Negative regulator nomenclature IDs` – list of negative regulators of the element.
N. `Interaction type` – for each listed regulator, in case it is known whether interaction is direct or indirect.
O. `Interaction mechanism` – for each known direct interaction, if the mechanism of interaction is known. Mechanisms that can be obtained from reading engines are listed in Table 3.
P. `Interaction score` – for each interaction, a confidence score obtained from reading.

The third group of fields in our representation includes **interaction-related provenance** information.

Q. `Reference paper IDs` – for each interaction, we list IDs of published papers that mention the interaction. This information is obtained directly from reading output.
R. `Sentences` – for each interaction, we list sentences describing the interaction. This information is obtained directly from reading output.

It is worth mentioning that this representation format can be converted into the SBML format to be used by different software tools and shared between different working groups. Additionally, the tabular format provides a readable interface that can be easily created or read by biologists, and generated or parsed by a machine.

## 4 From reading to model

We obtain outputs from three types of reading engines, namely REACH [2], RUBICON [26], and Leidos table reading (LTR)[27]. These reading engines provide output files with similar but not exactly the same format. In Table 3 we list the interaction mechanisms that can be obtained from these reading engines and in the following subsections, we outline the differences and advantages when working with all three reading

**Table 3.** Mechanisms recognized by the three reading engines.

| Reading Engine | Recognized Mechanisms |
| --- | --- |
| REACH [23] | Activation, Inhibition, Binding, Phosphorylation, Dephosphorylation, Ubiquitination, Acetylation, Methylation, Increase or Decrease Amount, Transcription, Translocation. |
| RUBICON [24] | Activation, Inhibition, Promotes, Signaling, Reduce, Induce, Supports, Attenuates, Stimulate, Antagonize, Synergize, Increase and Decrease Amount, Abrogates. |
| LTR [25] | Binding, Phosphorylation, Dephosphorylation, Isomerizations. |

Table 4. Converting REACH output for complexes into our modeling representation format.

| Column Name | Element | | | Positive Regulator | | Mech. Type | Paper ID | Evidence |
|---|---|---|---|---|---|---|---|---|
| | Name | Type | ID | Name | ID | | | |
| REACH Output | {FAK, PTP-PEST} | {Protein, Protein} | {Q05397, Q05209} | PIN1 | Q13526 | Binding | PMC 3272802 | PIN1 stimulates the binding of FAK to PTP-PEST |
| DySE Format Comp. 1 | FAK | Protein | Q05397 | PIN1 *AND* PTP-PEST | (Q13526, Q05209) | | PMC 3272802 | |
| DySE Format Comp. 2 | PTP-PEST | Protein | Q05209 | PIN1 *AND* FAK | (Q13526, Q05397) | | PMC 3272802 | |

engines.

### 4.1 Simple interaction translation

The first type of reading engine, REACH [2], can extract both direct and indirect interactions, as well as interaction mechanisms, where available. The simplest, and most common, reading outputs are those that include only a regulated element and a single regulator, each of them having one of the entity types listed in Table 1, and the interaction mechanism being one of the mechanisms described in Table 3. Such interactions have straight forward translation to our representation format, that is, they are translated into a single table row with some or all of the fields described in Section 3. Given that our modeling formalism accounts for positive and negative regulators, while reading engines can also output specific mechanisms where available in text, we assume in the translation that Phosphorylation, Acetylation, Increase Amount, and Methylation represent positive regulations, and Dephosphorylation, Ubiquitination, Decrease Amount, and Demethylation represent negative regulations. Additionally, we treat Transcription events as positive regulation.

### 4.2 Translation of translocation interaction

We translate translocation events (moving components from one cellular location to another) using the formalism described in [28]. This formalism requires including two separate model elements for the translocated component, one at each the original and the new location. Additionally, in the translocation type of interaction, translocation regulators can be listed.

### 4.3 Translation of complexes

Binding interaction mechanism represents formation of protein complexes in most cases. However, in order to include both individual proteins and complexes in which they participate within a single model, we defined rules for incorporating complexes listed in reading outputs into our model representation format.

A generic example is shown in Figure 3. If an element in the reading output file is a complex, we incorporate that output into our model representation format by creating a separate table row for each component of the protein complex, and change the regulation set as described in the example outlined in Figure 3. If the formation of complex AB is regulated by C, then we create two rows; one for element A, which is also positively regulated by F, and one for element B. The positive regulation rule for element A becomes (C AND B) OR F, while the positive regulation rule for element B becomes (C AND A). Additionally, if an element is regulated by a complex, we list all components of that complex as positive regulators for the element. In the example in Figure 3, the positive regulation rule for element D is (A AND B) because D is regulated by the complex AB. An example of how complexes are translated from reading output into our representation format is shown in Table 4.

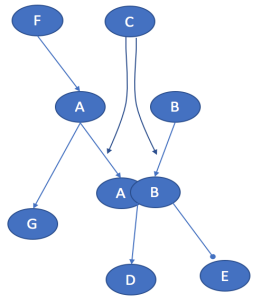

**Fig. 3**. Regulation of complexes.

### 4.4 Translation of nested interactions

REACH reading engine can also detect nested interactions, where some of the participants are interactions themselves. The following sub-sections show several examples of these interactions.

**Positive Regulation of Activation.** As shown in Figure 4a), REACH can find and output interactions where element A is activating element B, while element C is positively regulating the interaction between A and B. We also include in this example element D, a negative regulator of B. This means that C will activate B only when A is active. If A is inactive, only D will inhibit B, while C will not have any effect on B. The following is an example of the aforementioned situation that can occur in text, and is extracted by REACH as described above: "*In fact, RANKL induced phosphorylation of Akt was enhanced by the addition of TNF-alpha*". Here, RANKL is a positive regulator of Akt, and this activation is further regulated by TNF-alpha.

**Positive Regulation of Inhibition.** Figure 4b) illustrates an example of a nested interaction where A inhibits B, and C positively regulates this inhibition, which means that C will increase the inhibition of B by A, when A is active/high. We also include in this example element D, a positive regulator of B. If A is inactive/low, only D will activate B, and C will not have any effect on B. The following text represents an example sentence for such situation: "*This conclusion was supported by the finding that nilotinib also induced dephosphorylation of the BCR-ABL1 target CrkL*". Here, the inhibition of BCR-ABL1 by CrkL is enhanced by nilotinib.

**Negative Regulation of Activation.** The example in Figure 4c) shows that C negatively regulates the activation of B by A. So, if A is inactive/low, only D will activate B, and C will not have any effect on B. An example text for this situation is "*These data provide evidence that PDK1 negatively regulates TGF-β signaling through modulation of the direct interaction between the TGF-β receptor and Smad3 and -7*".

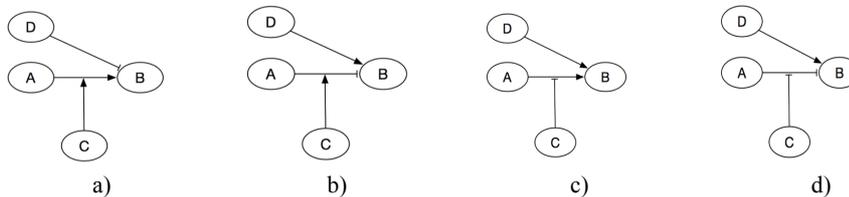

a)   b)   c)   d)

**Fig. 4.** Examples of nested interactions. a) Positive regulation of Activation interaction, b) Positive regulation of Inhibition interaction, c) Negative regulation of Activation interaction, d) Negative regulation of Inhibition interaction

**Negative Regulation of Inhibition.** Figure 4d) shows that C negatively regulates the inhibition of B by A. Therefore, if A is inactive/low, only D will activate B, and C will not have any effect on B.

### 4.5 Translation of direct and indirect interactions

RUBICON [26] provides two reading outputs, one for *direct interactions* and one for *indirect interactions*. For the indirect interactions, it creates a chain of elements that starts with the regulator and ends with the regulated element and includes the intermediate elements, also found in the read paper, forming a path from the regulator to the regulated elements.

The Direct output file contains direct interactions and uses two additional columns (compared to 4): Confidence and Tags. The Confidence column indicates how *confident* the reading engine is about the extracted interaction, and the values in this column can be LOW, MODERATE, and HIGH. The Tags column includes *epistemic tags* such as 'implication', 'method', 'hypothesis', 'result', 'goal', or 'fact'. Table 5 shows reading output examples from RUBICON for the direct and chain interactions. Due to space constraints, Table 5 does not include all the columns from the representation, as some columns are not filled in by RUBICON output.

The second reading output from RUBICON is the Chain file, which contains indirect interactions that form a path from the regulator to the regulated element. The Chain output contains a column called "Connection" and includes the name of the intermediate variable followed by its ID. So, if we have a path of the form A → B → C, element B will be included in the connection column.

### 4.6 Translation from table reading output

The third reading engine, LTR, performs table reading and generates reading output in the tabular format with some or all of the fields described in Section 3. The LTR output also contains information about Cell Line and Binding sites. Additionally, this output includes much more specific, connected information than those offered by RUBICON or REACH. Where RUBICON or REACH look at all the interactions listed in a paper, the nature of their search returns information on many different experiments and contexts. LTR is able to focus on one table at a time. As tables tend to describe a highly specific experiment about interacting components, such output can provide detailed information about parts of the network, which can be valuable in finding answers to specific questions. An example of an LTR output is shown in Table 6.

## 5 Matching reading and modeling

Due to the writing style in biology, reading engines often encounter texts that are hard to interpret even by human readers. In the following, we outline several situations where it is critical to correctly interpret interactions listed in reading outputs to enable accurate model expansion. When there are contradictions among reading outputs, or between reading output and an existing model, a feedback to reading can be generated in the form of new queries to guide further literature search and reading. Queries are designed using AND, OR and NOT to define more precisely the search space and also to remove papers that would describe information that is not relevant (e.g., focusing on different cell type).

**Table 5.** RUBICON output examples for both Direct and Chain.

| Column Name | Element | | Positive Regulator | | Mech. Type | Connection | Paper ID | Evidence | Confidence | Tags |
|---|---|---|---|---|---|---|---|---|---|---|
| | Name | ID | Name | ID | | | | | | |
| Direct | TNF alpha | P01375 | IL-2 | P60568 | induced | NA | PMC 149405 | In addition, cytokines including TNFalpha , TNF-beta and flt3 ligand were induced by IL-2 as detected by the arrays. | low | results |
| Chain | apoptosis | GO: 0006915 | imatinib | 5291 | enhances, induced | TRAIL, ID: P50591 | PMC 4896164 | Treatment with imatinib enhances TRAIL induced apoptosis | - | goal |

## 5.1 Protein families

Reading engines often come across entities that represent protein families instead of specific proteins. In such cases, there is no unique protein ID, instead either all IDs of proteins from that family need to be listed, or a unique protein family ID should be used. Since our goal is to automate the assembly of models from machine reading output, we need to be able to accurately treat such protein family entities in the reading output. There are several issues that can arise when protein families are outputs as interaction entities in reading output, described in the following example.

*Example 1*: Let us assume that either an existing model or previous reading output include an interaction that describes positive regulation of ERK1 by MEK1 (MEK1→ERK1), where both MEK1 and ERK1 are specific proteins that have unique IDs in protein databases. We list below other similar interactions that may be recognized by reading, and propose methods to resolve such situations.

**a.** Reading output MEK→ERK, where both MEK and ERK are listed as protein families. In order to incorporate both the original interaction and the new one within the same model, we can treat the new interaction as generalization. Furthermore, this is also an example of a situation where a feedback to reading engines can be created, to obtain more information about the interaction. For example, queries that could result from the scenario described here are:
- Search for other (non-MEK1) MEK family members and their interactions with ERK1;
- Search for other (non-ERK1) ERK family members and their interactions with MEK1;
- Search for other MEK (non-MEK1) and ERK (non-ERK1) family members, and their mutual interactions.

**b.** Reading output MEK1→ERK, where MEK1 is a protein and ERK is a protein family. In this case, the feedback to reading could be:
- Search for other ERK family members and their interactions with MEK1.

**c.** Reading output: MEK→ERK1, where MEK is a protein family and ERK1 is a protein. In this case, the feedback to reading could be:
- Search for other MEK family members and their interaction with ERK1.

**d.** Reading output: MEK→p38, MEK protein family activating protein p38. This case requires additional knowledge that would either already exist in the model or other reading outputs, or would need to be curated by a human expert. MEK3, and not MEK1, therefore, adding the original interaction (MEK1→ERK1) to the model, and then incorporating connection between MEK1 (as a member of MEK family) and p38 in the model would make it incorrect. The feedback to reading in this case could be:

- Search for interaction between MEK1 and p38 to confirm or disconfirm the interaction MEK→p38.

**Table 6.** LEIDOS output example illustrating the effects of the negative regulator (TiO$_2$) on two different molecules. As both sites affected by the negative regulator are serine residues, this provides additional context that the negative regulator might be a serine-specific.

| Element | | | Negative Regulator | | Cell Line | Organism | Paper ID | Evidence |
|---|---|---|---|---|---|---|---|---|
| Name | ID | Site | Name | ID | | | | |
| AKT1 | P31749 | S124 | TiO$_2$ | CHEBI: 32234 | HeLa | Human | PMC 3251015 | Resource3.xls.table.serial.txt |
| Gab2 | Q9UQC2 | S264 | TiO$_2$ | CHEBI: 32234 | HeLa | Human | PMC 3251015 | Resource4.xls.table.serial.txt |

### 5.2 Cell type

Often, the modeling goal is to include multiple cell types, for example, model of cancer microenvironment could include cancer cell and several types of immune cells. In such cases, it is important to know to which cell type to assign the interaction that is extracted from text by machine reading. When cell type is taken into account, the relationship between similar reading outputs, or between reading outputs and an existing model, can be interpreted in several ways.

*Example 2*: Let us assume that the machine reading output lists interaction A→B (A regulates B), but no information is given about cell type to which this interaction belongs. The model assembly step needs to decide to which cell to add this interaction, and therefore, different scenarios are possible, some of them described here:
- A is already listed in interactions in more than one cell type in the model;
- B is already listed in interactions in more than one cell type in the model;
- Neither A nor B is listed in other interactions;
- Both A and B are listed in interactions in exactly one cell type in the model (same or different).

The assembly step needs to either take into account previously defined assumptions (e.g., always add interactions to one predetermined cell type, or add interactions to all cell types, or skip the interaction that does not indicate cell type, etc.). Another approach is to request from readers to conduct additional search for evidence of cell type in the paper.

### 5.3 Cellular location

In some cases, it is important to know the location of elements participating in interactions. For example, translocation of element from one cellular location to another may take time, or it may be known that a particular element has effect on another element only in a specific location. In order to accurately model the interaction, the machine reading output should include the information about subcellular locations or extracellular space, the effect of location on interactions and on timing of cellular events (e.g., translocation).

*Example 3*: Let us assume that new reading output includes interaction A→B (A regulates B), but the interaction location is different from the one that exists in the current model. This can either be interpreted as a contradiction, or a feedback to reading can be generated in the form of a query to initiate literature search for further evidence of new interaction location. Additionally, the confidence obtained from reading can be compared with the confidence for the interaction in the model, to decide how to treat the reading output.

*Example 4*: Let us assume that an existing model includes interaction A→B (A positively regulates B) at a specific location, and reading output includes interaction A-|B (A negatively regulates B), but without location information. This can either be interpreted as a contradiction, or there can be a feedback to reading to search for further evidence of new interaction location. It is possible that the new interaction is observed at a different location, thus, the opposite regulation sign will not be interpreted as contradiction.

### 5.4 Contradicting interaction type

In the case of contradiction among individual reading outputs, or between new reading output and an existing model, a feedback to reading can be created to initiate new literature search.

*Example 5*: Let us assume that an existing model includes interaction A→B (A positively regulates B), while in reading output A-|B (A negatively regulates B). Assuming that the location information matches, there are several ways to in terpret this. This can be interpreted as a contradiction, or the new interaction may be indirect, forming a negative feed-forward loop with the one existing in the model. In this case, a feedback to reading can request search for further evidence for elements on a path between A and B.

### 5.5 Negative information

When it is well known that some interactions do not exist, such information is not stored in models. However, the reading output may include such interactions and the following example shows how such situation can be resolved.

*Example 6*: Let us assume that the previous reading output or an existing model includes interactions MEK1→ERK1 and MEK3→p38. There are several other reading outcomes that could occur:

**a.** New reading output includes interaction NOT (MEK3→ERK), where MEK3 is interpreted as a protein, and ERK is interpreted as a protein family. This is in agreement with the model, however, reading output that indicates that an interaction does not exist is not used to extend the model.

**b.** New reading output includes interaction NOT (MEK→ERK1), where MEK is interpreted as a protein family and ERK1 is interpreted as a protein. This new reading output

**c.** would contradict the model or other reading output, assuming that an interaction MEK1→ERK1 (from *Example 1*) already exists in the model or in other reading output. However, when taking into account the fact that MEK3 does not indeed regulate ERK1, such reading output could also be interpreted as corroboration. To resolve this, a search for further evidence in the paper indicating that the MEK from reading output is not MEK1 could be conducted.

## 6 Case study

To illustrate the utility of the translation from reading output to model representation format, we show an example of two queries, followed by a summary of reading results that we obtained, and the percentage of the results that we were able to use in the model.

### 6.1 Query 1: GAB2

The first query that we used is related to molecule GAB2. The original model does not contain GAB2 and we were interested in extending the model to incorporate GAB2. The query that we used is:

*GAB2 AND (phosphatidylinositol OR proliferation OR SHC1 OR PI-3 kinase OR PI3K OR PIK3 OR GRB2 OR PTPN11 OR 14-3-3 OR SFN OR YWHAH OR HCK OR AKT OR beta-catenin OR Calcineurin OR SERPINE1) NOT (Fc-epsilon receptor OR osteoclast OR mast cell)*

Note that GAB2 was identified in 1998 so the protein and gene have the same name and there is little confusion in the search. In Table 7, we show the number of papers returned by REACH and RUBICON reading engines using the GAB2 query above, the events extracted from all of the papers analyzed, and the unique extensions that were found by comparison to two existing models, Normal and Cancer.

### 6.2 Query 2: β-catenin

The second query that we used is related to molecule β-catenin. The original model does not contain β-catenin and we were interested in extending the model to incorporate this molecule. The query that we used is:

*(beta-catenin OR B-catenin OR β-catenin OR catenin beta-1 OR CTNNB1) AND (Wnt OR AXIN1 OR AXIN2 OR AXIN OR APC OR CSNK1A1 OR GSK3B OR TCF OR LEF OR TCF/LEF OR CDK2 OR PTPN6 OR CCEACAM1 OR insulin OR PML OR RANBP2 OR YAP1 OR GSK3 OR HSPB8 OR SERPINE1 OR AKT OR PTPN13 OR ACAP1 OR MST1R) NOT (neuroblasts OR neurogenesis OR anoikis OR cardiac OR EMT OR breast OR embryonic OR osteoblast OR synapse OR muscle OR renal)*

In this case, the β-catenin protein was identified in 1989 and the human gene in 1996 so the protein and gene have different names. However, using Greek letters in the name requires using various related terms in the query to increase the chance of capturing the right molecule in papers.

**Table 7.** Results from GAB2 query.

|                   | REACH | RUBICON |
|-------------------|-------|---------|
| Number of Papers  | 249   | 249     |
| Extracted Events  | 4800  | 2450    |
| Unique Extensions | 4618  | 3318    |

**Table 8:** Results from β-catenin query.

|                   | REACH | RUBICON |
|-------------------|-------|---------|
| Number of Papers  | 351   | 351     |
| Extracted Events  | 2809  | 1338    |
| Unique Extensions | 2532  | 1907    |

## 7 Conclusion

This paper describes a representation format that we created for the purpose of automating assembly of models from machine reading outputs. The proposed representation format allows for capturing biological interactions at the molecular level, and it can be easily used by both human experts and machines. By using this format, our automated framework rapidly assembles and validates executable models from big data in literature, with the runtimes and comprehensiveness not previously possible. Such formalized representation of research findings for the purpose of creating dynamic models will significantly speed up the process of collecting data from literature, and it will facilitate the reusability of existing scientific results, increase our knowledge and improve our understanding of biological systems. This, in turn, should lead to rapidly designing new disease treatments and effectively guiding future studies.

## References

1. Miskov-Zivanov, N. *Automation of Biological Model Learning, Design and Analysis*. in *Proceedings of the 25th edition on Great Lakes Symposium on VLSI*. 2015. ACM.
2. Valenzuela-Escárcega, M.A., et al., *A domain-independent rule-based framework for event extraction.* ACL-IJCNLP 2015, 2015. **127**.
3. Hucka, M., et al., *The systems biology markup language (SBML): a medium for representation and exchange of biochemical network models.* Bioinformatics, 2003. **19**(4): p. 524-531.
4. Droste, P., et al., *Visualizing multi-omics data in metabolic networks with the software Omix—a case study.* Biosystems, 2011. **105**(2): p. 154-161.
5. Büchel, F., et al., *Qualitative translation of relations from BioPAX to SBML qual.* Bioinformatics, 2012. **28**(20): p. 2648-2653.
6. Faeder, J.R., M.L. Blinov, and W.S. Hlavacek, *Rule-based modeling of biochemical systems with BioNetGen.* Systems biology, 2009: p. 113-167.
7. Hedengren, J.D., et al., *Nonlinear modeling, estimation and predictive control in APMonitor.* Computers & Chemical Engineering, 2014. **70**: p. 133-148.
8. Albert, R., *Scale-free networks in cell biology.* Journal of cell science, 2005. **118**(21): p. 4947-4957.
9. Pawson, T. and J.D. Scott, *Protein phosphorylation in signaling–50 years and counting.* Trends in biochemical sciences, 2005. **30**(6): p. 286-290.
10. Erwin, D.H. and E.H. Davidson, *The evolution of hierarchical gene regulatory networks.* Nature Reviews Genetics, 2009. **10**(2): p. 141-148.
11. Schuster, S., D.A. Fell, and T. Dandekar, *A general definition of metabolic pathways useful for systematic organization and analysis of complex metabolic networks.* Nature biotechnology, 2000. **18**(3): p. 326-332.
12. Schmitz, M.L., et al., *Signal integration, crosstalk mechanisms and networks in the function of inflammatory cytokines.* Biochimica et Biophysica Acta (BBA)-Molecular Cell Research, 2011. **1813**(12): p. 2165-2175.
13. Miskov-Zivanov, N., D. Marculescu, and J.R. Faeder. *Dynamic behavior of cell signaling networks: model design and analysis automation*. in *Proceedings of the 50th Annual Design Automation Conference*. 2013. ACM.
14. Sayed, K., et al., *DiSH simulator: Capturing dynamics of cellular signaling with heterogeneous knowledge* 2017.
15. UniProt. *UniProt Database*. Available from: http://www.uniprot.org/.
16. Pfam. *Pfam Database*. Available from: http://pfam.xfam.org/.
17. InterPro. *InterPro Database*. Available from: https://www.ebi.ac.uk/interpro/.
18. Bioentities, *Bioentities Database*.
19. PubChem. *PubChem Database*. Available from: https://pubchem.ncbi.nlm.nih.gov/.
20. HGNC. *Database of Human Gene Names*. Available from: http://www.genenames.org/.
21. GO. *Gene Ontology Database*. Available from: http://geneontology.org/page/go-database.
22. MeSH, *MeSH Database*.
23. REACH. *Reading and Assembling Contextual and Holistic Mechanisms from Text*. 2016; Available from: http://agathon.sista.arizona.edu:8080/odinweb/.
24. RUBICON, *Machine Reading Engine - Personal Communication*. 2016.
25. LTR, *Leidos Table Reader - Personal Communication*. 2016.
26. Burns, G.A., et al., *Automated detection of discourse segment and experimental types from the text of cancer pathway results sections.* Database, 2016. **2016**: p. baw122.
27. Sloate, S., et al., *Extracting Protein-Reaction Information from Tables of Unpredictable Format and Content in the Molecular Biology Literature*, in *Bioinformatics and Artificial Intelligence (BAI)*. 2016: New York, USA.
28. Sayed, K., C.A. Telmer, and N. Miskov-Zivanov. *Motif modeling for cell signaling networks*. in *Biomedical Engineering Conference (CIBEC), 2016 8th Cairo International*. 2016. IEEE.